\documentclass[aps,prb,preprint,superscriptaddress]{revtex4-2}
\usepackage{amsmath,amssymb}
\usepackage{graphicx}  
\usepackage{dcolumn}   
\usepackage{bm}        
 \usepackage{float}
\usepackage{amsfonts}
\usepackage{gensymb}
\usepackage{upgreek}
\usepackage{mathrsfs} 
 \usepackage{hyperref}

\hypersetup{
    colorlinks=true,
    linkcolor=blue,
    filecolor=magenta,      
    urlcolor=blue,
}
 
\usepackage{blindtext,tikz}
\usetikzlibrary{quantikz}
\usetikzlibrary{calc}

\begin{document}


\title{Simulation of Higher Dimensional Discrete Time Crystals on a Quantum Computer}


\author{Christopher Sims}
\affiliation{Elmore Family school of Electrical and Computer Engineering, Purdue University}
\email{Sims58@purdue.edu}


\date{\today}

\begin{abstract}
The study of topologically ordered states have given rise to a growing interest in symmetry protected states in quantum matter. Recently, this theory has been extended to quantum many body systems which demonstrate ordered states at low temperature. An example of this is the discrete time crystal (DTC) which has been demonstrated in a real quantum computer and in driven systems. These states are periodic in time and are protected to disorder to a certain extent. In general, DTC can be classified into two phases, the stable many body localization (MBL) state, and the disordered thermal state. This work demonstrates by generalizing DTC to 2 dimensions, there is an decrease in thermal noise and an increase in the operating range of the MBL range in the presence of disorder.
\end{abstract}

\maketitle

\section{Introduction}
An emerging topic in condensed matter physics is the engineering of new states of matter via floquet driving of systems . Floquet driving can create new hybrid topological states and ordered phases in matter which do not exist in normal systems \cite{Wang2013,Bukov2015,Harper2020,Titum2016}. Crystalline structures are examples of many body systems which can give rise to spatially ordered systems under periodic driving. In recent years, the study of discrete time crystals has gained considerable interest as a variation on floquet systems\cite{GomezLeon2013}.

Discrete time crystals (DTC) are periodically driven quantum many body systems which break time translation symmetry under certain driving conditions. These systems have been predicted theoretically and have been observed in periodically driven crystal systems \cite{Else2016,Keyserlingk2016,Sacha2017,Wilczek2012,Watanabe2015,Huse2013,Pekker2014}. DTC can be realized into two phases, the strongly disordered thermal phase which occured when there is low long range disorder and the many body-localization phase where eigenstates are observable over a long time period and have no dissipation with discrete spectra. Discrete time crystals have been realized experimentally in cold atom systems and in condensed matter systems with a periodic laser pulse\cite{Basko2006,Zhang2017,Choi2017,Rovny2018}. Building on previous work, these many body systems have been realized on real quantum computers and in quantum simulations \cite{Mi2021,Frey2022}. Furthermore, there has been a rising theoretical interest in studying DTC with longer periodic phases and systems which map to higher dimensions ($\geq$2)\cite{Kuros2021}. The study of time crystals which exist in higher dimensional lattice provides additional degrees of freedom in order to study time translational symmetry with multiple different phases in the same system. Recently, there has been an observation of a 2D discrete time crystal which was observed in a 1T-TaS$_2$ crystal system with periodic driving\cite{Gao2022,Yoshida2015,Lee2022}. In addition, there has been theoretical modeling of a magnetically ordered discrete time crystal which show that DTCs in two dimensions have a larger protection against thermal disorder in the one dimensional system \cite{Santini2022}. When this model is further extended to $N\geq2$ dimensions there is a further protection against thermal disorder with diminishing returns with higher dimensions \cite{Pizzi2021,Pizzi2021a}.

This works presents an extension of 1D discrete time crystals realized with quantum algorithms. With quantum simulation, it is demonstrated that 2D time crystals are more protected to disorder then 1D time crystals. In addition, this theoretical work can be extended into higher dimensional time crystals which can also be realized in quantum hardware.
\section{Discrete Time Crystals}
1D DTC can be described as a many-body Bloch-Floquet hamiltonian with phase flip and disorder terms \cite{Nandkishore2015,Abanin2019,Ponte2015,Lazarides2015,Bordia2017}.
\begin{equation}
\begin{aligned}
H = &  {} H_1  + H_2 + H_3\\
H_1 =  & g(1-\epsilon) \sum_i \sigma_i^y \\
H_2 =  & \sum_i J_{i,j} \sigma_i^x \sigma_j^x \\
H_3 =  & \sum_i D_{i} \sigma_i^x 
\end{aligned}
\label{BFH}
\end{equation}
Where $J_{i,j}$ is the Ising coupling,  $\Omega$ is the Rabi frequency, $\omega_R = \hbar \Delta k^2/(2M)$ is the recoil frequency, $b_{i,m}$ is the normal-mode matrix, and $\omega_m$ are the traverse mode frequencies \cite{Throckmorton2021,Alet2018,Chen2022,Hu2017,Kjaell2014,Hauke2015}.

 \begin{figure}[ht]
 \includegraphics[width=1.0\textwidth]{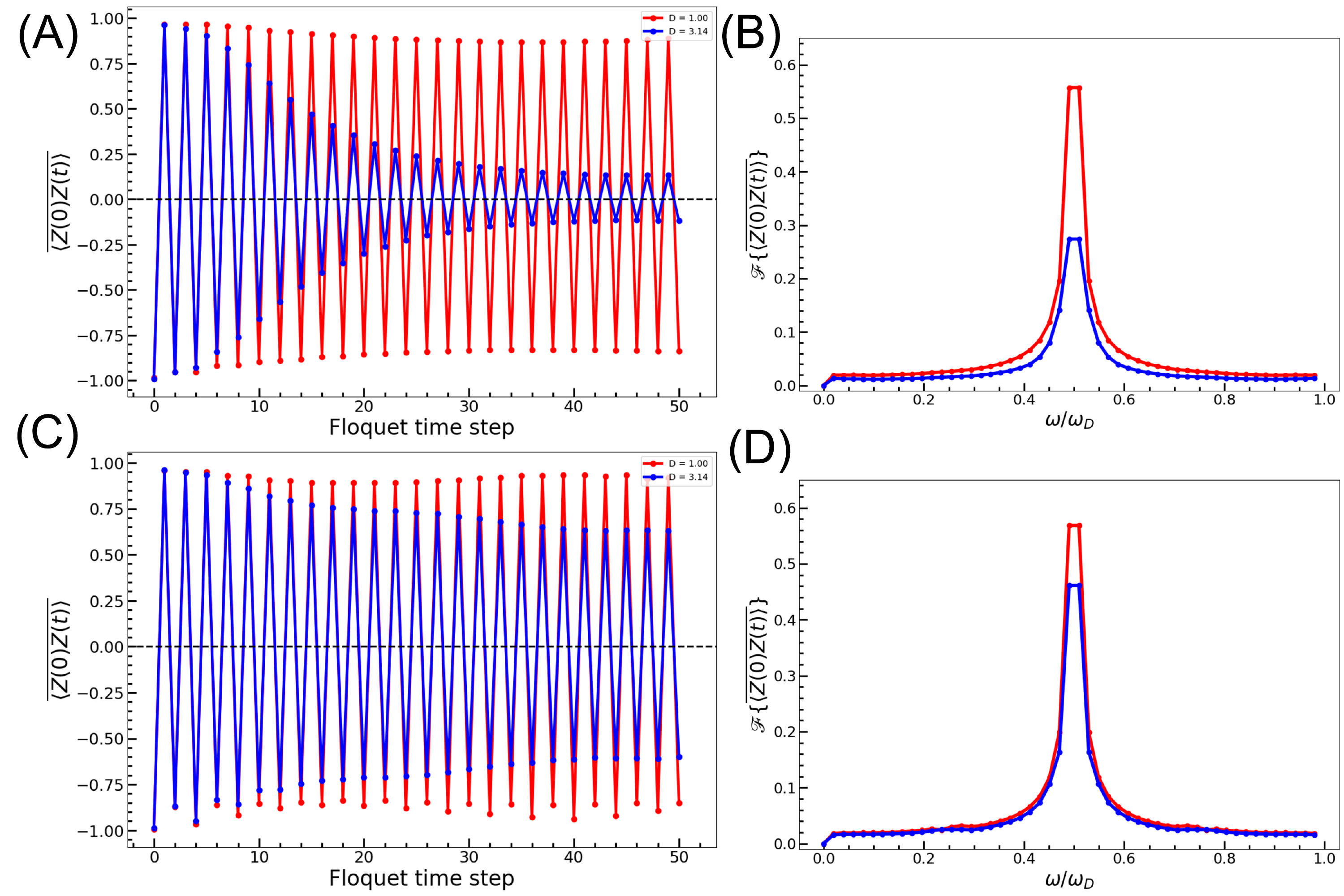}%
 \caption{\textbf{Autocorrelation:} The autocorrelation measurement averaged over all qubits $\overline{\langle Z(0)Zi(t)\rangle}$ with $g=0.9$ red denotes the low disorder system (D = 1) and blue denotes the high disorder system (D = 3.14) (A) 1D DTC system (B) the Fourier transform of the autocorrelation for the 1D system $\mathscr{F}\overline{\langle Z(0)Zi(t)\rangle}$ (C) the 2D DTC system with the same parameters as the 1D system (D) the Fourier transform of the autocorrelation for the 2D system $\mathscr{F}\overline{\langle Z(0)Zi(t)\rangle}$}
\label{AC}
 \end{figure}
The Ising Hamiltonian $H$ is a theoretical model in order to define time crystals in 2D. This model can be easily expanded into $N$-dimensions by generalizing the hopping Hamiltonians for their next nearest neighbor. When modeling a discrete time crystal in a quantum computing, this can be achieved by applying the gates which define the hopping terms to the next-nearest gates in the modeled system \cite{Dai2003,Inagaki2016,Leeuwen1975,BRUSH1967,Montroll1963,Onsager1944,Yang1952}.

When simulated and realized on quantum computing architecture, it is possible to see the DTC phase by analyzing the autocorrelation between the first time step in a qubit and its state after n time steps. $\overline{\langle Z_i(0)Z_i(t)\rangle}$. In the thermal regime of the DTC these states are expected to decay rapidly. However, in the MBL phase of the DTC these states are expected to remain at the same magnitude and oscillate between the two states. With modern day NISQ quantum computers, thermal noise is expected to depolarize the qubits over time in the quantum computer and cause a gradual decay of the states in the MBL phase of the DTC. For all simulations reported, the initial polarization is set to be fully polarized, which has been determined to have the most stable qubit states over time.

\section{1D vs. 2D DTC}
\begin{figure}[!ht]
\centering
 \includegraphics[width=1.0\textwidth]{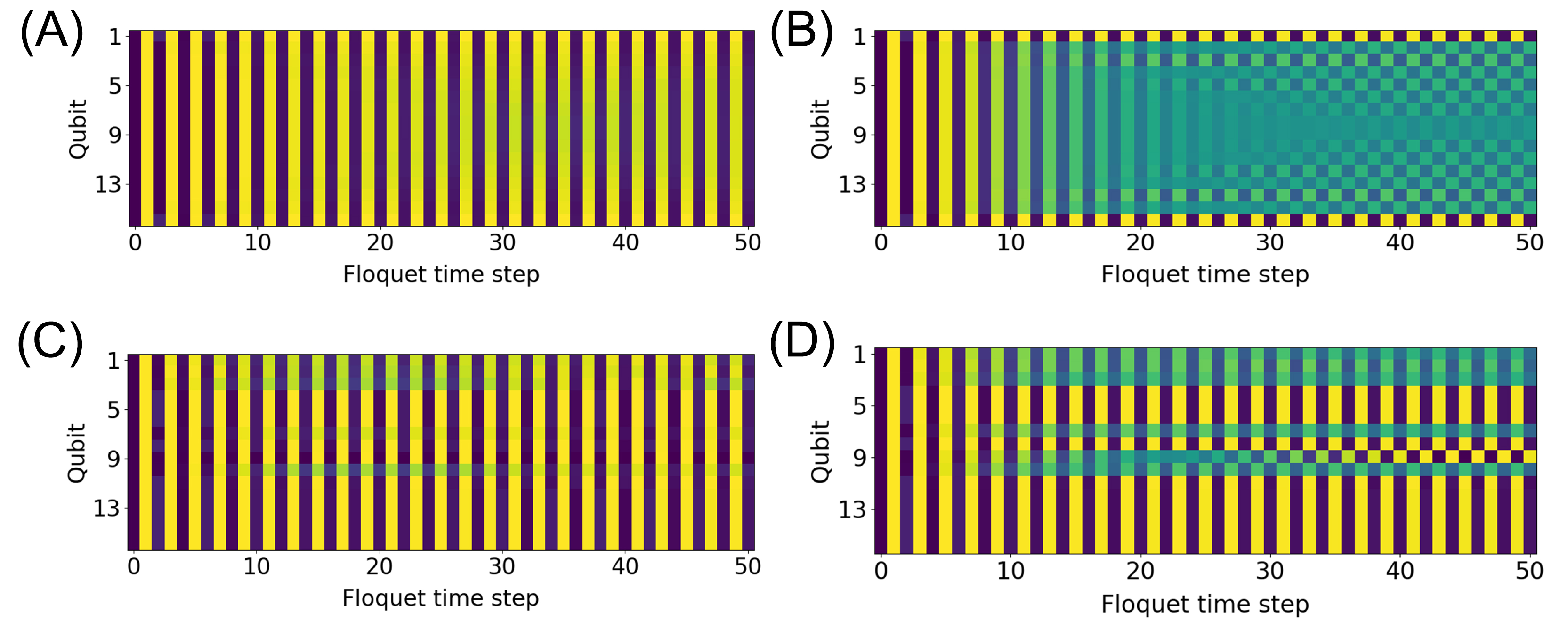}%
 \caption{\textbf{Polarization measurement:} The polarization measurement $\overline{A}$ = $\overline{\langle Z_i(0)Z_i(t)\rangle}$ of each qubit for each time step $t$ in the DTC state with spin interaction strength $g=0.9$ (A) 1D DTC with weak long range interaction strength D = $1$ (B) 1D DTC with strong long range interaction strength D = $\pi$ (C) 2D DTC with weak long range interaction strength D = $1$ (D) 2D DTC with strong long range interaction strength D = $\pi$}
\label{pol}
 \end{figure}

The 1D DTC is modeled with the Cirq quantum simulator for 50 floquet time steps with 16 qubits and $g=0.9$. In order to add noise into this simulation, a disorder parameter $D$ is added.  When $D=1$, [Fig \ref{AC}(A)] the autocorrelation $\overline{\langle Z(0)Z(t)\rangle}$ shows that since there is less disorder in the system and there is less loss of magnitude (red lines) and there is minimal depolarization in the system. Since $g = 0.9$, the system is in the strong MBL phase. Fourier transform analysis shows that there is strong order in the system  [Fig \ref{AC}(B)]. This can be further explored by analyzing the polarization of the qubits for the 1D DTC system with $D=1$ [Fig \ref{pol}(A)] where it can be seen that there is strong time crystal order in all qubits with correlated phases. When $D$ is increased to $\pi$ the disorder begins to take a larger effect on the total coherence of the system. The autocorrelation shows that [Fig \ref{AC}(A)] the system begins with a strong phase that decreases in average amplitude from fully polarized $\overline{\langle Z(0)Z(1)\rangle} = 1$ to disordered $\overline{\langle Z(0)Z(50)\rangle} = 0.25$. Fourier analysis shows a decrease in amplitude by $\sim\frac{1}{2}$ from 0.6 to 0.3 for the normalized amplitude [Fig \ref{AC}(B)]. The disordered evolution of the 1D DTC can be examined with a polarization measurement which shows a an ordered state until $t=10$ then a transition to a disordered state with increasing time steps [Fig \ref{pol}(B)].

	The 2D discrete time crystal is an expanded form of the 1D discrete time crystal. There are now an increased amount of spin interaction terms ($R_{ZZ}$ gates) between nearest-neighbors when the qubits are arranged in a ``virtual'' 2D crystal which corresponds to the 2D grid layout of the quantum computer. For comparison, the same disorder parameter is used for the 2D case as the 1D DTC simulation. 

WIP
In the event of minimal disorder, the 2D DTC [Fig \ref{pol}(C,D), red] has similar performance to the 1D DTC [Fig \ref{pol}(A,B), red]. However, when D = $\pi$ the 2D DTC [Fig \ref{pol}(C,D), blue] has far less loss $\approx 4.3$ than the 1D DTC system $\approx 2.6$ this shows that there is minimal loss of polarization in the 2D DTC when compared to a lower order system. This demonstrates that there is an increased protection against disorder as the order of the discrete time crystal system is increased. Additionally, When D = 1, [Fig \ref{pol}(C),red], there is a minimal loss of polarization after $t$ time steps and there is a recovery of the polarization of the states as the 2D time crystal develops in time (time steps $t = 25-50$, whereas the 1D time crystal has a steadily decaying polarization state in the low disorder state.

\section{Discussion}
The increase protection from disorder can be explained by a stronger interaction between qubits because there are more nearest neighbor qubits interacting with each other. The Ising interaction $J_{i,j}$ scales as $J_0/|i-j|^\alpha$. This means that the nearest neighbor bits contribute the most to the protection against disorder, and the corner or edge states are the most prone to disorder. By reformulating the qubits from a 1D chain to a 2D lattice, it is possible to have more interaction terms with nearest neighbors. For the first nearest neighbor term, this goes from 2 for the 1D case to 4 for the 2D case with a square lattice. This effect can be scaled to $N$ dimensions with an increase in nearest neighbor protections to the qubits.

\begin{figure}[!h]
\centering
\begin{quantikz}
\lstick{\ket{x_0}} & \gate{R_y(\pi)} \gategroup[4,steps=7,style={dashed,
rounded corners,fill=blue!20, inner xsep=2pt},
background]{{\sc $t$ time steps}} & \ctrl{1} &\qw&\ctrl{2}&\qw&\ctrl{3}&\gate{XZ(D)} &\meter{}\\
\lstick{\ket{x_1}} & \gate{R_y(\pi)} &  \gate{R_{ZZ}(\theta)} & \ctrl{1} &\qw&\ctrl{2}&\qw&\gate{XZ(D)}  &\meter{}\\
\lstick{\ket{x_2}} & \gate{R_y(\pi)}  &\ctrl{1} &  \gate{R_{ZZ}(\theta)}&\gate{R_{ZZ}(\theta)}& \qw&\qw  &\gate{XZ(D)} &\meter{}\\
\lstick{\ket{x_3}} & \gate{R_y(\pi)}  & \gate{R_{ZZ}(\theta)} & \qw&\qw&\gate{R_{ZZ}(\theta)}&\gate{R_{ZZ}(\theta)}&\gate{XZ(D)} &\meter{}
\end{quantikz}
\caption{\textbf{2D DTC circuit}: An example of a 2D DTC crystal algorithm with 4 qubits implemented, this results in a 2$\times$2 square with interacting qubits. This algorithm is repeated for $t$ Bloch-Floquet time steps and then measured at the end. $\vec{x}$ corresponds to the initial state, which can be of N\'eel, polarized, or random order.}
\label{circ}
\end{figure}
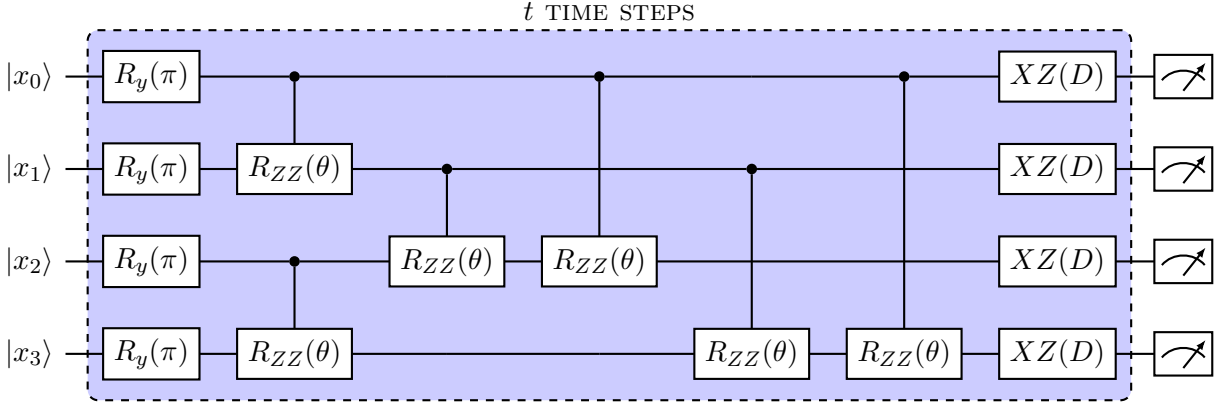
\section{Methods}
\subsection{Simulation}
All calculations were performed in python v 3.9.1 with the Google quantum AI Cirq package v1.0.0. GPU acceleration is enabled with the Nvidia cuQuantum 22.07.1 SDK. Within the Cirq library the FSIM gates are used with constant perimeters $\theta = \pi$, $\phi = \pi$, $\alpha = \pi$, $\beta = \pi$. Qubits were simulated with a grid layout similar to Google quantum machines. All qubits are typically measured in the $\hat{Z}$ basis in a quantum machine. However, for this simulation, qubits are measured using CIRQ's vector state simulator, the polarization measurement is then done in the $\hat{Z}$ basis in order to get the qubit rotation with respect to the $\hat{Z}$ axis.
\subsection{Edge State}
When computing the average of the qubit states, all qubits are taken into consideration. Edge qubits and their effects are considered to be part of the system and are integral to the results.
\subsection{Circuits}
The time crystal algorithm can be split into 3 main stages, the rotations, the mixing parameters and finally the long range interactions, these stages correspond to $H_1$, $H_2$, $H_3$ of equation- \ref{BFH}. The bloch floquet time step can be translated into a quantum algorithm by 3 simple gates. The first stage of gates are $R_y$ single gate rotations, for this work, all gates are rotated by $\pi$ since this is the most stable, and optimal rotation for a real quantum algorithm. Next is the ``spin'' interactions, which can be modeled ideally with an 2 qubit $R_{ZZ}$ gate. This gate belongs to the more generalized FSIM gates which has been used in previous works to fine tune the circuit, however, the $R_{ZZ}$ variation is used in this work. In order for form a 2D state the $R_{ZZ}$ gates are applied to the nearest neighbors (NN) with corresponding qubits theortically arranged in a square. Since Google quantum computers are arranged in a square pattern, this algorithm can be realized with an NN algorithm. Finally is the phased XZ gate with the exponent $\alpha = 0$ resulting in a rotation about the X axis with a phase determined by $Z(D)$.

\section{Conclusion}
In conclusion, a method is developed to create two dimensional time crystals utilizing quantum computing architecture \& quantum algorithms. It is found that 2D time crystals are more robust then 1D time crystals and are more impervious to disorder in their respective system. The method of adding additional dimensions how a diminishing return in reducing noise in the system for a compounding increase in the number of qubits to define a system with the same edge lengths. In addition, $N\geq3$ dimensional time crystals require qubits to have adjacent qubits for fast, low noise swaps, this is unlikely to exist for quantum hardware which is confined to 2D or 3D in future architecture. Nevertheless, there is an interest to show how higher dimensional time crystals behave on real quantum hardware. 

\section{Code availability}
Code will be made publicly available upon full publication.


 \begin{acknowledgments}
C.S. acknowledges the generous support from the GEM Fellowship and the Purdue Engineering ASIRE Fellowship.
 
 Correspondence and requests for materials should be addressed to C.S.
 (Email: Sims58@Purdue.edu)
\end{acknowledgments}


%

\end{document}